\documentclass[a4paper, amsfonts, amssymb, amsmath, reprint,onecolumn]{revtex4-2}

\usepackage[utf8]{inputenc}
\usepackage{graphicx}
\usepackage{natbib}
\usepackage{gensymb}
\usepackage{xcolor}
\usepackage{color}
\usepackage{cancel}
\usepackage{amsmath}
\usepackage{bm}
\usepackage{array}
\usepackage{amsfonts}
\usepackage{amssymb}
\usepackage{epstopdf}
\usepackage{subfiles}
\usepackage{soul}
\usepackage{float}

\newcommand{\largebullet}{\large$\bullet$}

\usepackage[utf8]{inputenc} 
\usepackage[T1]{fontenc}    
\usepackage{hyperref}       
\usepackage{url}            
\usepackage{booktabs}       
\usepackage{amsfonts}       
\usepackage{nicefrac}       
\usepackage{microtype}      
\usepackage{graphicx}
\usepackage{natbib}
\usepackage{siunitx}

\begin{document}

\title{Coalescence of viscous blisters under an elastic sheet}

\author{W-E. Khatla}
\author{L. Duchemin}
\email{laurent.duchemin@espci.fr}
\author{A. Eddi}
\author{E. Reyssat}

\affiliation{
PMMH, CNRS, ESPCI Paris, Université PSL,\\ 
Sorbonne Université, Université Paris Cité,\\ 
F-75005, Paris, France
}
\date{\today}

\begin{abstract}
We study the coalescence of identical viscous blisters beneath an elastic sheet both experimentally and numerically. Using a time-resolved synthetic schlieren technique, we measure the evolution of the thickness field of the merging blisters and more specifically the dynamics of the coalescence region. To explain this dynamics, we develop a one-dimensional model based on the lubrication approximation, from which we derive scalings to predict the growth of the coalescence neck. We also numerically solve the full non-linear equation to assess the theory and compare to experiments. Our model illustrates that, at short times, the dynamics of coalescence is mainly controlled by the bending of the elastic sheet, leading to a relationship between the coalescence speed and the radius of curvature of the interface at the coalescence neck.
\end{abstract}

\keywords{Coalescence \and Lubrication theory \and Thin film \and Fluid-structure interaction \and Elasticity}

\maketitle
\section{Introduction}

{The spreading of a viscous fluid on a solid substrate occurs at a diversity of scales in laboratory, industrial or natural phenomena. Droplets for inkjet printing technologies are typically 10 \textmu m in radius and spread within timescales as low as a millisecond \citep{Lohse2022}. Dermo-epidermal suction blisters are submillimetric liquid pockets, form within hours, and are used as a diagnosis tool for some bullous diseases \citep{Willsteed1991}.   
At the geological scale, the flow of lava results in the formation of a number of characteristic elongated structures whose final shape depends on the interplay of gravity, viscous stresses and strongly time-dependent rheology resulting from cooling and crusting processes \citep{Griffiths2000}. Among them are laccoliths, shallow structures with a typical thickness of order 100 m and width spanning typically 10 km, which form in 100 to 1000 years \citep{Bunger2011,Michaut2011,Mathieu2008}.\\
One key ingredient in the dynamics of these systems is the boundary condition at the surface of the liquid. While the spreading dynamics of droplets under the effect of surface tension has been studied extensively \citep{Bonn2009}, liquid is often covered by a skin or a crust whose solid character strongly influences the shape and dynamics of the underlying fluid. It has for instance been shown that the growth of blisters under quasi two-dimensional thin sheets gives rises to wrinkling instabilities characteristic of slender membranes under compression \citep{Pandey2025}. The displacement of a high viscosity liquid by a lesser viscous fluid in a Hele-Shaw geometry exhibits the classical Saffman and Taylor instability which is strongly affected when confinement is provided by a flexible membrane \citep{Pihler2012}. A number of recent experimental and theoretical studies describe such quasi-parallel viscous flows through model systems that couple fluid flow in the lubrication approximation to the mechanics of the covering elastic or plastic crust \citep{Lister2013,Hewitt2015,Peng2020,Saeter2023,Ball2023}. \\
Besides spreading, coalescence is another generic and practically important event in the life of a sessile drop. It has been explored in many configurations and regimes \citep{Eggers2025}. Specifically, the growth of the height and width of the liquid bridge forming in the  coalescence of wetting viscous droplets on a solid have been shown to follow scaling laws in time \citep{Ristenpart2006,Hernandez2012,Hack2020,Kaneelil2022}.\\ 
In the present paper, we focus on the onset of coalescence of two liquid blisters covered by an elastic membrane. A recent work combines experiments, lubrication theory and numerical simulations to address the coalescence dynamics of two identical drops underneath a thin elastic sheet \cite{Saeter2024}. We propose to revisit and discuss the results of this letter through experimental results done in a different parameter range, scaling models and numerical computations. We first present the experimental setup and geometric quantities measured. We then derive a scaling theory that links the upward velocity of the coalescing neck to its local radius of curvature. Finally, we present an idealized numerical study that confirms and discusses the short-time self-similar solution predicted by \citep{Saeter2024}.
}

\section{Experiments}

\begin{figure}[h!]
	\centering
	\includegraphics[width = .7\textwidth]{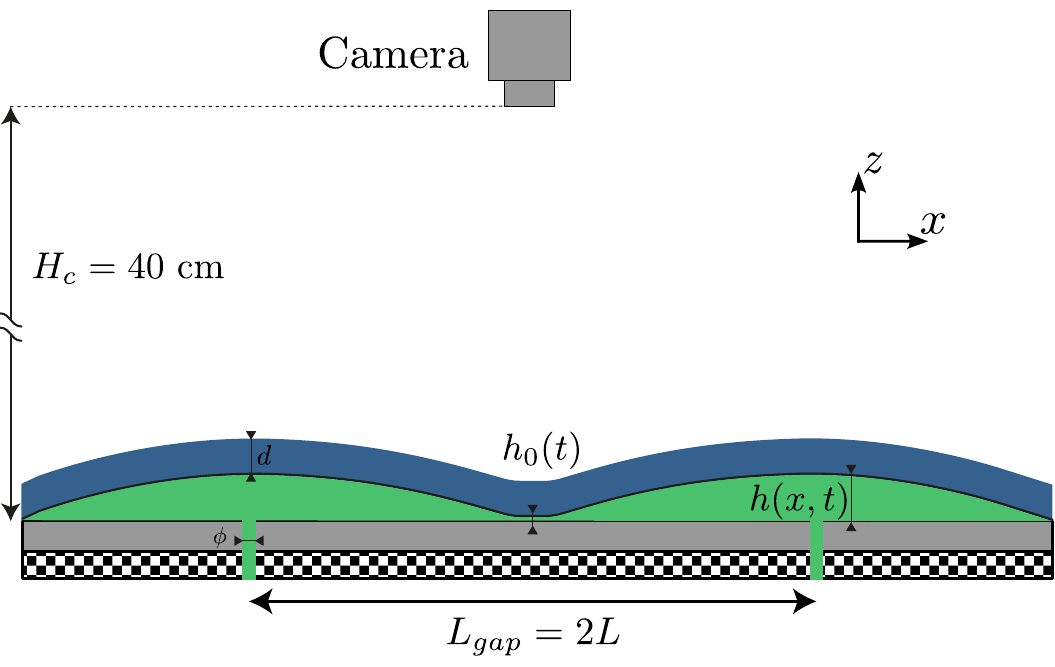}
	\caption{Sketch of the experimental setup. Sunflower oil is fed under an elastic membrane (thickness $d$) through two holes (diameter $\phi \simeq 1$ mm) made in an underlying transparent plate, forming two identical liquid blisters of typical thickness $h \sim 0.1-1$ mm separated by a distance $L_{gap} \sim 3$ cm. A checkerboard pattern is placed under the plate and imaged from the top through the whole setup. Apparent deformations of the pattern enable one to reconstruct the liquid thickness field $h(x,t)$.
}
	\label{fig:setup}
\end{figure}

{Two identical liquid blisters are formed by injecting sunflower oil (viscosity $\mu=5\times 10^{-2}$ Pa.s) between a planar PMMA plate and an elastic sheet (figure \ref{fig:setup}). The blisters are fed at constant flow rate $Q=0.05-12$ \textmu L.s$^{-1}$ through the PMMA substrate using two channels of diameter $\phi=0.7$ mm separated by a distance $L_{gap}\in [1.5, 3, 6]$ cm. Prior to the injection, the substrate is cleaned using ethanol, dried, and covered with the elastic sheet. The latter is made of PDMS (Silex \textit{Superclear}) with Young's modulus $E\simeq 1.6$ MPa and Poisson ratio $\nu \simeq 0.5$ \cite{domino2020artificial,Ledoudic2025}. We control the bending stiffness $B=Ed^3/12(1-\nu^2)$ by using membranes with different thicknesses $d \in [0.2, 0.8, 1.6]$ mm. Injection is stopped before the blisters get into contact. The total injected volume $V$ is then fixed and spreading goes on spontaneously beneath the elastic sheet until both liquid pockets meet and start coalescing. The values of $V$ and $Q$ are used to control the geometry of the membrane at the onset of coalescence.\\
The dynamic evolution of the thickness $h(x,t)$ of the liquid layer is extracted using a digital schlieren technique, \textit{i.e.} from the apparent deformations of a checkerboard pattern placed under the PMMA substrate and imaged from above through the different layers of the setup. A Basler camera with $2048\times 2048$ pixels is placed at a distance $H_c \simeq 40$ cm above the PMMA plate. The field of view is a square region of $10 \times 10$ cm$^2$, corresponding to a horizontal spatial resolution around 50 \textmu m/pixel. Images are acquired at 10 to 20 frames/second. The apparent displacement field of the reference pattern is fed into a Fast Checkerboard Demodulation (FCD) algorithm which results in the reconstruction of the deformed sheet height field $h(x,y,t)$ \citep{Moisy2009, Wildeman}. Calibration prefactors have been checked by integrating the measured height field during the constant flow rate injection phase of the experiment and ensuring that the resulting volume indeed grows linearly in time: $V=Qt$. The checkerboard tiles are 0.5 mm accross, which yields a vertical resolution of the method estimated at 10 \textmu m. We therefore obtain a complete cartography of the blisters height in time. An example of this reconstructed field is shown in figure \ref{figure2}a at the early times of the coalescence processes. In this particular example, $V=0.5$ mL, $L_{gap}=3$ cm, $d=0.2$ mm.

First, we identify the peaks of the blisters and analyze a uni-dimensional cut through the $(xz)$ plane that contains these two characteristic points. We thus measure the temporal evolution of the profile heights in-between these peaks during the total duration of the experiment. An overview of this procedure is presented in figure \ref{figure2}b where we can clearly notice the edges of the blisters advancing until they meet. Then, the coalescence process begins and results in the rise of the center point and the complete merging of the drops. We distinguish two phases in the evolution of the coalescence point height. First, a rapid rise ($\Delta t \sim 30 \si{s}$), followed by stabilization in the form of a height plateau corresponding to around $60\%$ of the initial maximum pre-coalescence blister height. 

\begin{figure}[t]
\centering
\includegraphics[width = .9 \textwidth]{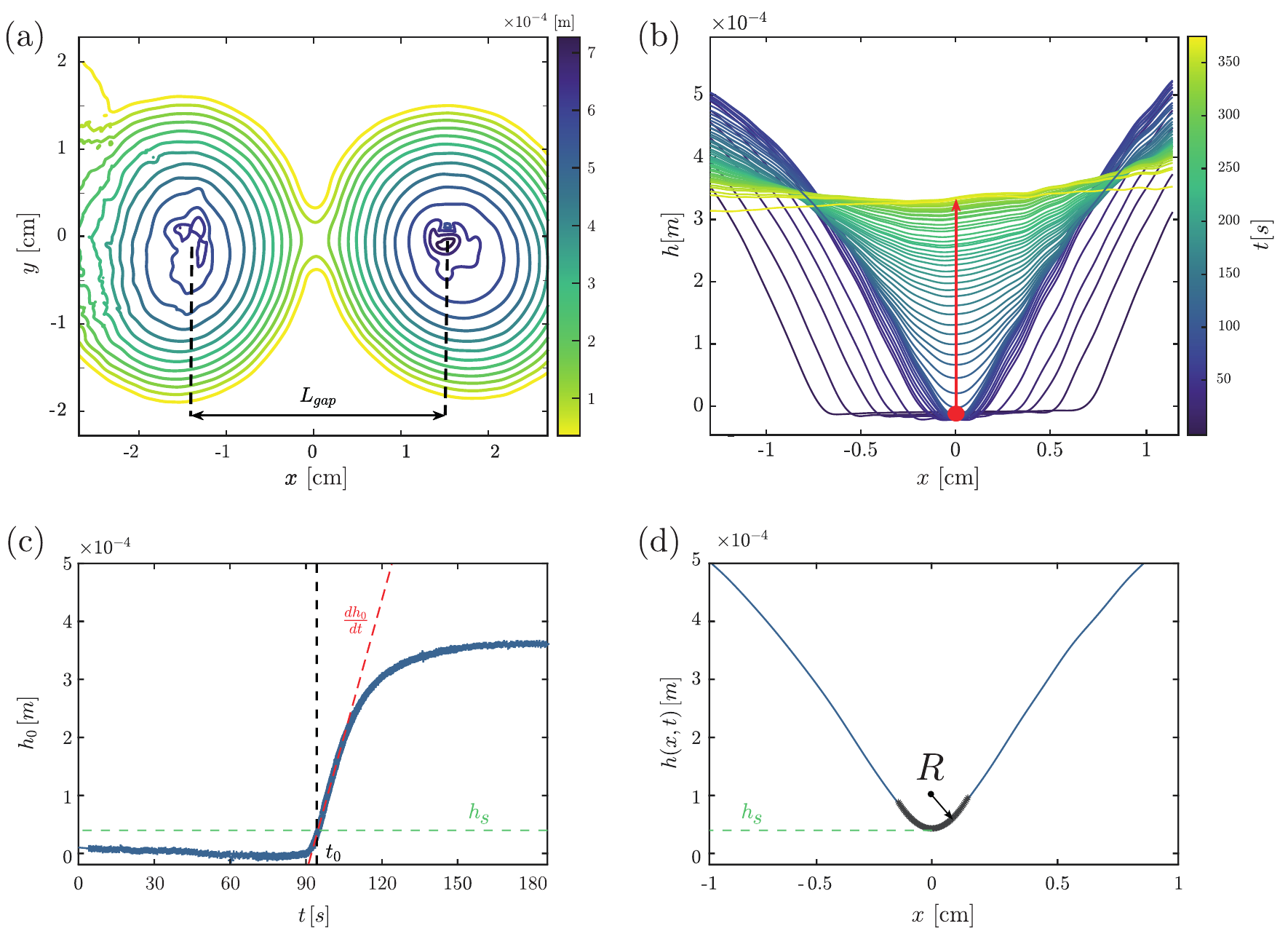}
\caption{ (a) Reconstruction of the height profile $h(x,y)$ at $t = t_0 + 2 \si{s}$. The contours of the blisters are clearly visible and the spacing between their respective centers corresponds to the imposed
inter-intrusion distance $L_{\text{gap}}$. Here,  $V=0.5$ mL, $L_{gap}=3$ cm and $d=0.2$ mm. (b) Height profiles between the centers of the injected blisters. The contact point position $x_0$ is highlighted as a red dot. Color gradient indicates increasing times. (c) Temporal evolution of the coalescence point height $h_0 (t)$. On the linear part, we measure a coalescence speed noted $dh_0/dt (t)$ when $h_0$ reaches $h_s = 40~\mu$m. (d) Identification of the height profile $h(x,t_0)$ and determination of the radius of curvature $R$ of a small region around the minimum }
\label{figure2}
\end{figure}

The first quantity extracted is the height $h_0(t)$ of the coalescence point. A typical result is shown in figure \ref{figure2}c, where we clearly observe the two phases discussed before. 
We measure the coalescence speed $dh_0/dt$ at a common bridge height $h_s = 40~\mu$m for all experiments. The value of $h_s$ is chosen to be three times the vertical resolution of the optical reconstruction method, a height at which we are sure that the coalescence process has begun (see figure \ref{figure2}c). For each experiment, we also extract the local radius of curvature $R$, obtained by fitting a second-order polynomial with a least square algorithm. For a typical test case the procedure is exhibited in figure \ref{figure2}d. 
Moreover, we measure the coalescence speed $dh_0/dt$ 
{at the coalescence time $t_0$} using a linear regression of $h_0(t)$.
We performed experiments with three different values of the thickness $d$ of the elastic sheets and plotted the relationship between $dh_0/dt$ and $R$ in logarithmic coordinates as illustrated in figure \ref{fig:loipuissance} (left). This clearly exhibits a power-law scaling which we propose to investigate. The captured slope appears constant through all values of the thickness $d$, {whereas the prefactor depends on $d$}.


\section{Scaling theory} 

In order to model the coalescence process, we use the lubrication theory to describe the evolution of the local thickness $h(x,y,t)$ of the liquid film under the joint influence of gravity and of the bending elasticity of the membrane. The thin-film equation reads:
\begin{equation}
    \frac{\partial h}{\partial t} = \frac{1}{12\mu}\nabla \cdot \left(h^3 \nabla 
    (\rho g h + B \nabla^4 h) 
    \right),
    \label{lubrication}
\end{equation}
where $B = Ed^3/12(1-\nu^2)$ is the bending modulus of the elastic membrane, $\rho$ is the density of the liquid, $g$ is the acceleration of gravity  
and $\nabla = (\partial_x,\partial_y)$ is the 2D gradient operator. 
Since, in our experiments, the vertical deflection of the elastic sheet is small compared to its thickness, we neglect the tension in the sheet with respect to bending stresses \cite{Lister2013}. 
Indeed, the fluid pressure induced by bending and tension scale respectively like $B h_{xxxx} \sim E d^3 h /\lambda^4$ and $T h_{xx} \sim E d (\sqrt{h^2+\lambda^2}-\lambda)/\lambda \times h/\lambda^2 \sim E d h^3 / \lambda^4$, where $\lambda$ is a typical horizontal deformation scale. Hence, bending dominates over tension if $B h_{xxxx} \gg T h_{xx}$, {\it i.e.} $d \gg h$. 

We are interested in the coalescence region between the two bumps (figure \ref{fig:setup}), around the point where $\partial h/\partial x = 0$. In the experiments, the transverse curvature along the $y$-axis is typically 10 times smaller than in the $xz$-plane. Therefore, neglecting the transverse curvature, we start with a 1D approximation of equation \eqref{lubrication}:
\begin{equation}
    \frac{\partial h}{\partial t} = \frac{1}{12\mu} \frac{\partial}{\partial x} \left(h^3 
    \left[ 
    \rho g \frac{\partial h}{\partial x}
    +
    B \frac{\partial^5 h}{\partial x^5}
    \right]
    \right)
    \label{eq:h}
\end{equation}

Around the minimum height $h_0$, the natural horizontal and vertical scales are related through the local curvature of the interface:
\begin{equation}
\frac{1}{R} \sim \frac{h_0}{\delta x^2}.
    \label{eq:xscale}
\end{equation}

In other words, in the vicinity of the coalescence point, the thickness of the film increases by $h_0$ for a horizontal displacement $\delta x \sim \sqrt{Rh_0}$.
Using these local scales $\delta x$ and $h_0$, we estimate the ratio of the gravity and bending terms in equation \eqref{eq:h}, which is extremely small in the experiments: $4.4 \times 10^{-4} \leq \rho g h_0^2 R^2/B \leq 2.6 \times 10^{-3}$. Therefore, the bending term dominates and equation \eqref{eq:h} reduces to:
\begin{equation}
    \frac{\partial h}{\partial t} = \frac{B}{12\mu} \frac{\partial}{\partial x} \left(h^3 
    \frac{\partial^5 h}{\partial x^5}
    \right)
    \label{eq:h_bending}
\end{equation}

Expecting the coalescence process to depend on the local scales, we look for a self-similar solution of equation \eqref{eq:h_bending} in the form:
\begin{equation}
    h(x,t) = h_0(t) H(X),
    \label{eq:ansatz}
\end{equation}
where $X=x/\sqrt{R h_0}$. 
Plugging this ansatz into equation \eqref{eq:h_bending} leads to:
\begin{equation}
    \frac{d h_0}{dt} H(X)
    - \frac{h_0 X}{2} 
    \left(
    \frac{1}{R} \frac{dR}{dt} + \frac{1}{h_0} \frac{dh_0}{dt}
    \right)
    \frac{d H}{d X}
    =
    \frac{h_0}{R^3} \frac{B}{12\mu} \frac{d}{dX}\left( H^3 \frac{d^5H}{dX^5} \right)
    \label{eq:selfsim}
\end{equation}
In reference \cite{Saeter2024}, the authors look for a self-similar solution in the form $h(x,t)=f(t) H(x/g(t))$, where $f(t)$ and $g(t)$ are related to each other through a matching condition for $X\to \infty$, where the curvature $\kappa$ is constant and given by the quasi-static blister solution: $f(t)=\kappa g^2(t)$. 
Interestingly, in their formulation, the curvature at the coalescence point is also constant in time, since $\partial^2h/\partial x^2(0,t) = f(t)/g^2(t) H_{XX}(0)=\kappa H_{XX}(0)$. 

The difference with our approach is that we allow, through the scalings, the curvature at the coalescence point to evolve in time, starting from an initial finite value $1/R_0$. 
The theory presented in \cite{Saeter2024} is therefore valid as long as one can neglect $R(t)-R_0$ with respect to $R_0$. 

Initially, the radius of curvature has a finite value $R_0$. Therefore we can look for $R(t)$ as:
$$
R(t)=R_0 + \varepsilon R_1(t),
$$
where $\varepsilon \ll 1$ and equation  \eqref{eq:selfsim} reads, at leading-order in $\varepsilon$:
\begin{equation}
    \frac{d h_0}{dt} H(X)
    - \frac{X}{2} 
    \frac{dh_0}{dt}
    \frac{d H}{d X}
    =
    \frac{h_0(t)}{R_0^3} \frac{B}{12\mu} \frac{d}{dX}\left( H^3 \frac{d^5H}{dX^5} \right)
    \label{eq:selfsim0}
\end{equation}
At $X=0$, given that $H(0)=1$ and $dH/dX(0)=0$, we find:
\begin{equation}
    \frac{d h_0}{dt}
    =h_0(t)
    \frac{B}{12\mu R_0^3} \frac{d^6H}{dX^6}(0)
    \label{eq:selfsim1}
\end{equation}
which leads to a solution for $h_0(t)$ consistent with the short-time solution proposed in \cite{Saeter2024}:
\begin{equation}
    h_0(t) = h_0(0) e^{\alpha B t/12 \mu R_0^3},
\end{equation}
where $\alpha = d^6 H/d X^6(0) = \mathcal{O}(1)$. 
In order to assess this scaling, we compare the measured coalescence speed $dh_0/dt$ with the predicted law:
\begin{equation}
    \frac{d h_0}{dt}=\alpha \frac{B h_0}{12 \mu R_0^3},
    \label{scaling_v}
\end{equation}
where $\alpha=1$ gives a very good comparison with the experiments. 
Figure \ref{fig:loipuissance}(a) shows the coalescence speed $dh_0/dt$ measured when $h_0=h_s=40 \; \mu$m, as a function of the radius of curvature $R$, for three different membrane thicknesses : $d=0.2$, $0.8$ and $1.6$ mm. The three dashed lines correspond to equation \eqref{scaling_v} for $\alpha=1$. 
Rescaling the coalescence speed by the bending modulus $B=E d^3/12(1-\nu^2)$ allows to collapse all theses points on a master curve, as seen in figure \ref{fig:loipuissance} (right), confirming the linearity in $B$, that could have been anticipated from equation \eqref{eq:h_bending}.

\begin{figure}[t]
\centering
\includegraphics{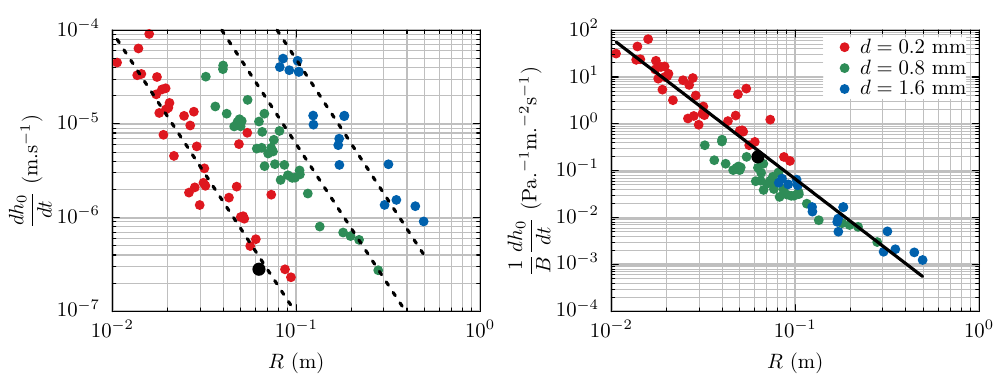}
\caption{{ \bf Left~:} Speed $dh_0/dt$ of the elevation of the coalescence point 
as a function of the radius of curvature $R$ on a log-log scale, for three different membranes thickness 
(({\color[rgb]{0.8666666666666667,0.09411764705882353,0.12156862745098039}{\largebullet}}) $d=0.2$ mm, 
({\color[rgb]{0.169117647058824,0.511029411764706,0.319852941176471}{\largebullet}}) $d=0.8$ mm, 
({\color[rgb]{0,0.3764705882352941,0.6784313725490196}{\largebullet}}) $d=1.6$ mm). {\bf Right~:} $dh_0/dt$ normalized by the bending modulus $B$ plotted as a function of $R$ on a log-log scale. The dashed and continuous straight lines correspond to equation \eqref{scaling_v} with $\alpha=1$. The black dot corresponds to the numerical computation presented in section \ref{sec:num}, using $L = 1.5$ cm, $d = 0.2$ mm, $E = 1.6 \times 10^6$ Pa, and $\mu = 0.05$ Pa.s..
}
\label{fig:loipuissance}
\end{figure}

\section{Numerical computation}
\label{sec:num}

In order to solve equation \eqref{eq:h_bending} numerically, we first make it dimensionless 
using a length-scale $L=L_{gap}/2$, ($x^\star=x/L$, $h^\star=h/L$) and a timescale $12\mu L^3/B$ ($t^\star=B t/12\mu L^3$), leading to:
\begin{equation}
    \frac{\partial h^\star}{\partial t^\star} = \frac{\partial}{\partial x^\star} \left(h^{\star3} 
    \frac{\partial^5 h^\star}{\partial x^{\star5}}
    \right).
    \label{eq:h_bending_adim}
\end{equation}
We impose a zero flux and symmetry boundary conditions at $x^\star=0$ and $x^\star=1$, {\it i.e.} 
$q=h^{\star3}\partial^5h^\star/\partial x^{\star5}=\partial h^\star/\partial x^\star=\partial^3h^\star/\partial x^{\star3}=0$. 

Owing to the fact that the dynamics is extremely sensitive to initial conditions (Cf. equation \eqref{scaling_v}), and that the flow is 3D in the experiments, it is difficult to make a quantitative comparison between experiments and numerics. Instead, we design an idealized polynomial initial condition, with a zero derivative in both $x^\star=0$ and $1$:
$$
h^\star(x^\star,0) = h^\star(0,0) + (h_m^\star-h^\star(0,0)) \left( 2 x^{\star2} -  x^{\star4} \right),
$$
such that $h^\star(1,0) = h_m^\star$. 
$h^\star(0,0)$ and $h_m^\star$ are chosen to match the values of the experiment presented in figure \ref{figure2}, {\it i.e.} $h^\star(0,0)=h(0,0)/L = 10^{-5} / 0.015 \simeq 6.67 \times 10^{-4}$  and $h_m^\star = h_m/L = 7 \times 10^{-4} / 0.015 \simeq 4.67 \times 10^{-2}$.
This initial condition would correspond to a static solution of equation \eqref{eq:h_bending_adim} if the other boundary conditions were satisfied, {\it i.e.} $q=\partial^3h^\star/\partial x^{\star3}=0$. 
We numerically integrate equation \eqref{eq:h_bending_adim}, using a semi-implicit finite differences code, second-order in space and time \citep{duchemin2020}. 
The time step is continuously adapted to ensure that the relative error on $h$ remains small at all times. 

\begin{figure}[t]
\centering
\includegraphics[width = \textwidth]{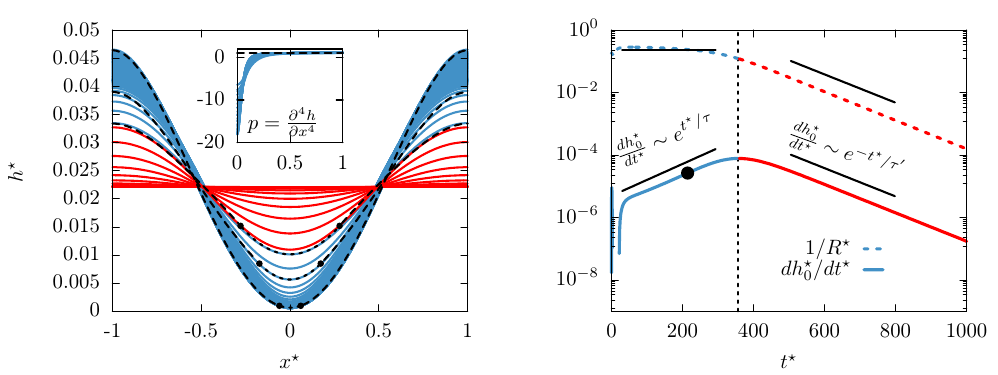}
\caption{{\bf Left~:} Successive profiles of the numerical solution of equation \eqref{eq:h_bending_adim}, relaxing towards a uniform height $\bar{h} \simeq 2.21\times10^{-2}$. The dotted and dashed curves are theoretical composite approximate solutions of equation \eqref{eq:h_bending_adim}. Inset~: Pressure field during the self-similar evolution. The horizontal dashed line corresponds to $p=1$.
{\bf Right~:} $dh^\star_0/dt^\star$ and $1/R^\star$ plotted in a log-lin scale as a function of time. 
The vertical dotted line in the right figure $(t^\star \simeq 356.5)$ separates the self-similar growth on the left (blue profiles), and the exponential relaxation toward $\bar{h}$ (red profiles).
}
\label{fig:simulation}
\end{figure}

Figure \ref{fig:simulation} shows the successive profiles of the interface (left) and the evolution of $dh^\star_0/dt^\star$ and $1/R^\star(t^\star)$ in a log-lin scale (right). The two figures share the same color rules: blue: $t^\star \le 356.5$, red: $t^\star \ge 356.5$. Crossing the vertical dotted line at $t^\star$, we observe that the coalescence velocity $dh_0^\star/dt^\star$ undergoes an exponential growth followed by an exponential decay. 

At short times ($t^\star \le 356.5$), we expect the self-similar evolution \eqref{eq:ansatz} to hold as long as $R^\star(t^\star)-R^\star_0 \ll R^\star_0$. If this condition is satisfied, the dimensionless version of equation \eqref{scaling_v} should hold:
$$
\frac{dh^\star_0}{dt^\star} = \alpha \frac{B h_0}{12 \mu R_0^3} \frac{12\mu L^2}{B} 
= \alpha \frac{h^\star_0}{R^{\star3}_0}
\equiv \alpha \frac{h^\star_0}{\tau}
$$
As long as $R^\star$ can be approximated by a constant, $dh^\star_0/dt^\star$ should evolve according to $\exp(\alpha t^\star / R^{\star3}_0)$, with $\alpha \simeq 1$. This is confirmed by the two black lines corresponding to a fitted timescale $\tau = R^{\star3}_0 = 4.4^3$, and to the constant value $1/R^\star_0$ of the curvature. The fact that $1/R^\star$ is indeed constant and close to $1/R^\star_0$ confirms the consistency of the self-similar solution. 
Moreover, in order to compare to the experiments, we measure the numerical velocity $dh^\star_0/dt^\star$ when $h(0,t)$ reaches $40 \; \mu$m, {\it i.e.} when $h^\star(0,t^\star) = 40 \times 10^{-6} / 0.015 \simeq 2.67 \times 10^{-3}$. This data is shown as a black dot in figure \ref{fig:simulation} and in figure \ref{fig:loipuissance}, when going back to dimensional quantities, using $L = 1.5$ cm, $d = 0.2$ mm, $E = 1.6 \times 10^6$ Pa, and $\mu = 0.05$ Pa.s. This measure is in good agreement with the experimental results, confirming the consistency between the numerics and the scaling law \eqref{scaling_v}. 

In addition to this local self-similar evolution, we can approximate the global dynamics by patching a leading-order expansion of the self-similar solution $h^\star \simeq h^\star_0 + x^{\star2}/2R^\star$ to a static outer solution $h_{out}(x,t)$. Following the scaling of equation  \eqref{eq:xscale}, we define the patching point as $\left[ x^\star_p,\, h^\star(x^\star_p)\right] =\left[ \sqrt{R_0^\star h_0^\star},\, 3h^\star_0/2\right]$. These composite approximations are shown in dotted and dashed lines in figure \ref{fig:simulation}, for three different values of $h^\star_0$ ($6.67\times10^{-4}, 5.67\times10^{-3}$ and $10^{-2}$): the dotted curves correspond to the approximate self-similar function $h^\star_0 + x^{\star2}/2R^\star$, and the dashed curves show a fourth-order polynomial, which describes the static region. The coefficients of this polynomial are obtained thanks to the following conditions: the pressure $p=\partial ^4 h^\star/\partial x^{\star 4}$ outside the self-similar region is found to be approximately constant, equal to $1$ (Inset in figure \ref{fig:simulation}), the volume below the full curve (dotted and dashed) must equal the initial value, the slope in $x^\star=1$ vanishes, the vertical position and slope at the patching point $\left[ x^\star_p,\, h^\star(x^\star_p)\right]$ must be the same as for the inner solution. The agreement between this composite solution and the full numerical solution is very good, showing that the dynamics occurs mostly in the region of coalescence, and that the outer shape of the membrane is mainly a static shape (constant pressure) connected to the lower region. 

At long times, the relaxation towards a flat membrane should follow a linear theory. 
In this case, $h^\star$ can be sought for in the form $h^\star(x^\star,t^\star) = \bar{h} - \delta(t^\star) \cos \pi x^\star$, 
where $\bar{h}$ is the dimensionless equilibrium average thickness and $\delta \ll \bar{h}$. Plugging this expression into equation \eqref{eq:h_bending_adim} and linearizing in the perturbation gives:
$$
-\frac{d \delta(t^\star)}{dt^\star} = \bar{h}^3 \pi^6 \delta(t^\star),
$$
which has the solution:
$$
\delta(t^\star) = \delta(0) e^{-t^\star/\tau'},
$$
where $\tau' = 1/\left(\bar{h}^3\pi^6\right)$. It follows that the coalescence speed $dh^\star_0/dt^\star$ and the curvature in $x^\star=0$ read:
$$
\frac{dh^\star_0}{dt^\star} \sim e^{-t^\star/\tau'}, 
\qquad 
\frac{1}{R^\star(t^\star)} = \frac{\partial^2 h^\star}{\partial x^{\star2}}(0,t^\star) \sim e^{-t^\star/\tau'}.
$$
On the right of the vertical dotted line, the dynamics indeed follows the linearized behavior $\exp(-t^\star/\tau')$, both for $dh^\star_0/dt^\star$ and $1/R^\star$.

\section{Conclusion}

In this study, we have conducted careful experiments of the coalescence of two thin blisters. The dynamics of the connecting bridge is understood in the pure bending regime thanks to the experiments, a scaling theory and 1D numerical simulations. The vertical velocity of the bridge is found to scale like $1/R^3$, where $R$ is the local radius of curvature of the membrane. A self-similar solution to the 1D idealized flow is consistent with a previous publication on the same subject \cite{Saeter2024} about the short-time dynamics. Contrary to \cite{Saeter2024} who assume a constant radius of curvature in time, our approach leaves its value free and the timescale for the exponential growth of the liquid bridge is shown to be consistent with the constant value of $R$.  We have extended this short-time dynamics by studying its transition to an exponential relaxation towards a flat membrane. In addition, we have shown that the local self-similar solution can be patched to a static outside shape, proving that the dynamics is mainly governed locally by the coalescence region. 

Among the promising outcomes of this study is definitely the extension to a full 2D numerical simulation ($z=h(x,y,t)$), that would allow for a quantitative comparison of the long-time dynamics with the experiments. Moreover, increasing the initial height of the bumps would require the inclusion of the tension in the membrane, that has been neglected in this study. This physical ingredient is more challenging to include in the numerics, since it requires a lagrangian tracking of the membrane. Including the tension would allow us not only to solve the long-time dynamics, but also to describe the appearance of wrinkles around the bumps, a consequence of an orthoradial buckling instability. Finally, the growth, spreading and coalescence dynamics of such blisters most likely depend on the adhesion of the membrane to the solid surface at the periphery of the bumps. This adhesion can be taken into account either by defining a binding energy, or by assuming the presence of a thin precursor film, that can also be treated thanks to lubrication theory.  

We thank Gregory Kozyreff for useful discussions and a careful proofreading.


\end{document}